\documentclass[twocolumn,english,prl,showpacs,preprintnumbers,superscriptaddress]{revtex4-2}
\usepackage[utf8]{inputenc}
\usepackage{color}
\usepackage{amsmath,amssymb}
\usepackage{amssymb}
\usepackage{graphicx}
\usepackage{textcomp}
\usepackage{dcolumn}
\usepackage{bm}
\usepackage[right]{eurosym}
\usepackage{float}
\usepackage[english]{babel}
\usepackage{blindtext}
\usepackage{babel}
\usepackage[normalem]{ulem}
\begin{document}
\title{Efficient Indirect Interatomic Coulombic Decay Induced by Photoelectron Impact Excitation in Large He Nanodroplets}

\author{L. Ben Ltaief}
\affiliation{Department of Physics and Astronomy, Aarhus University, 8000 Aarhus C, Denmark}
\author{K. Sishodia}
\affiliation{Quantum Center for Diamond and Department of Physics, Indian Institute of Technology, Madras, Chennai 600036, India}
\author{S. Mandal}
\affiliation{Indian Institute of Science Education and Research, Pune 411008, India}
\author{S. De}
\affiliation{Quantum Center for Diamond and Department of Physics, Indian Institute of Technology, Madras, Chennai 600036, India}
\author{S. R. Krishnan}
\affiliation{Quantum Center for Diamond and Department of Physics, Indian Institute of Technology, Madras, Chennai 600036, India}
\author{C. Medina}
\affiliation{Institute of Physics, University of Freiburg, 79104 Freiburg, Germany}
\author{N. Pal}
\affiliation{Elettra-Sincrotrone Trieste, 34149 Basovizza, Trieste, Italy}
\author{R. Richter}
\affiliation{Elettra-Sincrotrone Trieste, 34149 Basovizza, Trieste, Italy}
\author{T. Fennel}
\affiliation{Institute for Physics, University of Rostock, 18051 Rostock, Germany}
\author{M. Mudrich}
\affiliation{Department of Physics and Astronomy, Aarhus University, 8000 Aarhus C, Denmark}
\email{ltaief@phys.au.dk and mudrich@phys.au.dk}
\date{June 2022}

\begin{abstract}
Ionization of matter by energetic radiation generally causes complex secondary reactions which are hard to decipher. Using large helium nanodroplets irradiated by XUV photons, we show that the full chain of processes ensuing primary photoionization can be tracked in detail by means of high-resolution electron spectroscopy. We find that elastic and inelastic scattering of photoelectrons efficiently induces interatomic Coulombic decay (ICD) in the droplets. This type of indirect ICD even becomes the dominant process of electron emission in nearly the entire XUV range in large droplets with radius $\gtrsim40~$nm. Indirect ICD processes induced by electron scattering likely play an important role in other condensed phase systems exposed to ionizing radiation as well, including biological matter. 
\end{abstract}

\maketitle

Ionization of matter by extreme ultraviolet (XUV) and x-ray radiation involves cascades of secondary processes including electron scattering, intramolecular relaxation, and intermolecular transfer of charge and energy. These cascades comprise both ultrafast electronic processes and nuclear motion, spanning time scales of femtoseconds up to nanoseconds and beyond. Unravelling the details of such ionization cascades generally is a formidable task due to the high complexity of interactions and the resulting congestion of experimental spectra~\cite{Stumpf:2016}. However, understanding and even controlling ionization mechanisms in condensed phase systems, in particular those producing low-energy electrons that cause radiation damage in biological matter~\cite{Alizadeh:2015,Boudaiffa:2000}, is crucial for improving radiotherapies~\cite{sanche2016interaction}. 
Here we present a comprehensive study of the ionization mechanisms of a model condensed-phase system -- nanometer-sized droplets of helium (He) -- by XUV radiation. Owing to the simple electronic structure of the He atom and the peculiar quantum-fluid properties of He nanodroplets, we are able to resolve all primary and secondary ionization processes up to the level of involved excited quantum states and charge states of the products. In particular, we find that both elastic and inelastic electron collisions at He atoms in the droplets efficiently induce further ionizations by Interatomic Coulombic Decay (ICD)~\cite{Cederbaum:1997,Kuleff:2010}. For large He droplets with radius $\gtrsim40$~nm, electron impact-induced ICD even becomes the dominant channel of electron and ion emission. Variants of ICD have been observed both for pure He droplets and for heterogeneous droplets doped with foreign species. In those experiments, He droplets were resonantly excited or simultaneously ionized and excited by XUV radiation~\cite{Peterka:2006,Wang:2008,Ovcharenko:2014,Shcherbinin:2017,Wiegandt:2019,Buchta:2013,Ltaief:2019,Ovcharenko:2020,LaForge:2021}. ICD-like processes have also been detected as minor channels in strong-field ionized He droplets and rare-gas clusters where electron-ion recombination lead to the population of excited atomic states~\cite{SchuetteCorrelated:2015,oelze:2017,Kelbg:2019,Kelbg:2020}. 

The indirect ICD process reported here is not induced directly by XUV radiation but by electrons emitted by primary photoionization, subsequently interacting with He atoms of the droplet. We observe it for both pure and doped He droplets in a broad range of photon energies $h\nu>44$~eV up to the soft x-ray range. While this indirect ICD process is particularly well discernable in He nanodroplets, we believe that it plays an important role in all condensed-phase media subjected to ionizing radiation. In particular molecular systems such as liquid water tend to form electronically excited states (excitons) by electron-ion recombination which subsequently decay by ICD. While excitons have been found to diffuse over large distances in water ice~\cite{Petrik:2003}, their characteristics and mobility remain poorely understood~\cite{garrett2004role}.

For the analysis of secondary processes leading to ICD we use
He nanodroplets and performed two experiments at the Gasphase beamline at the Synchrotron radiation facility Elettra Trieste, Italy. The setups are schematically depicted in Fig.~1. In the first one [Fig.~1 a)], a mobile He droplet apparatus was attached to an electron-ion coincidence velocity-map imaging (VMI) spectrometer~\cite{OKeeffe:2011}. This arrangement has been described before~\cite{Buchta:2013,BuchtaJCP:2013}. In short, a continuous beam of He nanodroplets containing $10^4$ up to $\sim 10^8$ He atoms per droplet is generated by expanding He out of a cryogenic nozzle at a temperature ranging from 16 down to 8~K and 50~bar He backing pressure. A mechanical chopper is used for discriminating all the spectra shown here from background.
\begin{figure}[!h]
	\center	\includegraphics[width=0.9\columnwidth]{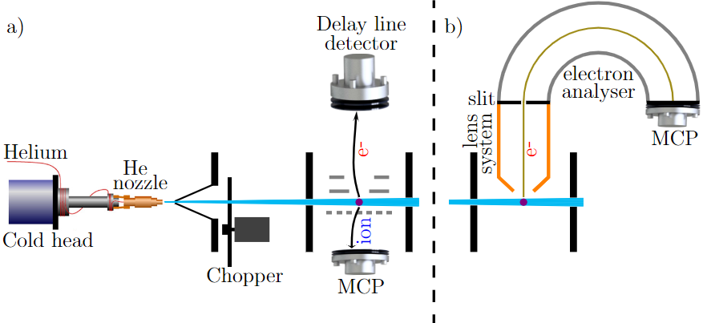}\caption{\label{fig1} Sketch of the experimental setups used in this work. a) He nanodroplet beam source and coincidence velocity-map imaging (VMI) spectrometer. b) Hemispherical electron analyzer.}
\end{figure}

In the detector chamber further downstream, the He droplet beam intersects the XUV beam with flux $\Phi\approx10^{14}$~s$^{-1}$cm$^{-2}$ in the VMI spectrometer at right angles. Electron spectra are inferred from electron images by Abel inversion~\cite{Dick:2014}. This setup allows us to simultaneously measure spectra of all emitted electrons and of electrons emitted in coincidence with specific fragment ions in a wide energy range. In the second arrangement [Fig.~1 b)], an hemispherical electron analyzer (VG-220i)
mounted at the magic angle is used to measure high-resolution electron spectra.

\begin{figure}[!h]
	\center
    \includegraphics[width=0.8\columnwidth]{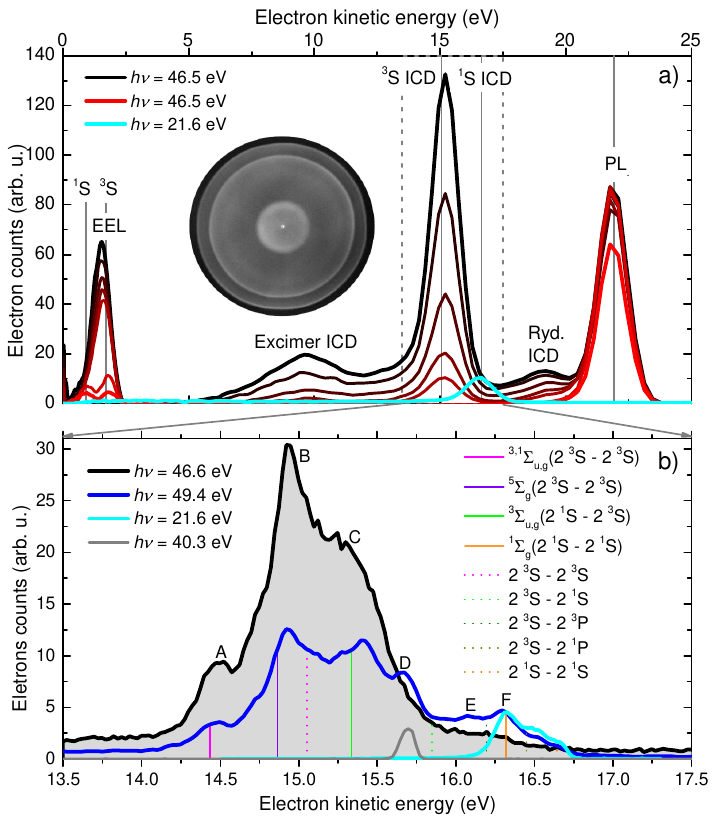}\caption{\label{fig2} a) Electron spectra of variable size measured at various $h\nu$'s and inferred from VMI's as that shown in inset ($h\nu=46.5$~eV, $R=50$~nm). Red and black lines are spectra of He droplets of radii $R=5$, 7, 20, 25, 35, 40 and 75~nm (red to black). The cyan line is a reference spectrum at $h\nu=21.6~$eV and $R=50$~nm. The vertical solid lines indicate expected electron energies based on atomic He levels. b) High-resolution analyzer spectra measured around the main ICD peak for $R=50$~nm and at various $h\nu$'s. The dashed and solid vertical lines indicate characteristic electron energies for ICD involving two He$^*$$-$He$^*$ formed in their $^{3}$S and $^{1}$S atomic states and two He$^*$$-$He$^*$ that form $\Sigma$ and $\Pi$ molecular states, respectively. The gray line shows photoelectrons detected at $h\nu=40.3$~eV used for calibration purposes. All spectra are normalized to the photon flux and to the $h\nu$-dependent absorption cross section of He.}
\end{figure}

The present study focuses on the photon-energy range $h\nu\gtrsim 44$~eV in which He droplets are directly photoionized and the emitted electron can further excite or ionize a He atom in the droplet by inelastic collision. As example, at $h\nu=46.5$~eV [see Fig.~2 a)], electron spectra contain both a photoline at electron energy $E_e=h\nu - E_i=21.9$~eV, where $E_i=24.6$~eV is the ionization energy of He, and an electron energy-loss (EEL) peak at $E_e - E_*\approx 2$~eV, where $E_*$ is the energy of the impact-excited state~\cite{Shcherbinin:2019}. The colored lines indicate various He droplet sizes in the range of 5 to 75~nm (red to black). The EEL peaks in the kinetic energy range $0.2$-$2.5$~eV are due to photoelectron-helium ($e$-He) impact excitation. EEL spectra revealing the states that are impact excited at different $h\nu$'s are shown in Fig.~1 of the Supplemental Material (SM). 

Besides these common features present for all droplet sizes, more features appear and grow in the electron spectra when the droplet radius is increased from $R=5$~nm [red line in Fig.~2 a)] to $\gtrsim20$~nm (black line). The salient new feature is a peak around 15~eV which even surpasses the photoline for large droplets with $R\gtrsim40$~nm. This peak remains nearly constant in energy for all $h\nu>45~$eV (see SM Fig.~2). Its energy is close to that previously observed in experiments where He droplets were resonantly excited by intense XUV pulses from a free-electron laser (FEL)~\cite{Ovcharenko:2020,LaForge:2021}. There, multiply excited He droplets decayed by ICD according to the reaction $\mathrm{He}^* + \mathrm{He}^* \xrightarrow{} \mathrm{He}^+ + \mathrm{He} + e_\mathrm{ICD}$
occurring in He droplets~\cite{Kuleff:2010}. He$^*$ stands for an excited He atom in the lowest optically accessible excited 1s2s\,$^1$S state. In fact, when we tune $h\nu$ to the strongest resonance of He droplets at $h\nu=21.6$~eV (1s2p\,$^1$P state) in the present synchrotron experiment, we also detect ICD out of the 1s2s\,$^1$S state leading to emission of electrons with kinetic energy around 16.5~eV (cyan line in Fig.~2). Although the photon flux is lower than in FEL experiments by orders of magnitude, large He droplets can still be multiply excited owing to their large resonant absorption cross section (25~Mb for each He atom in the droplet, see SM Sec.~IV and~\cite{BuchtaJCP:2013}). At $h\nu=46.5$~eV, the absorption cross section of He is only 2.3~Mb~\cite{Samson:2002} and excitation of He requires secondary $e$-He inelastic collisions which occurs with probability $<1$. Nevertheless, we detect ICD electrons at $h\nu=46.5$~eV with factor 20 higher rate than at $h\nu=21.6$~eV when taking the $h\nu$-dependent absorption cross section into account. Therefore, we will argue that a different secondary process facilitates the formation of pairs of He$^*$'s as precursors of ICD: Consecutive $e$-He impact excitation and elastic scattering of the slowed electron leads to recombination with its parent ion as depicted in Fig.~3~a). This one-photon process is even more efficient than ICD induced by multiple photon absorption by the droplets under the present conditions. 

To get more detailed insight into this new ICD process we use the electron analyzer to measure ICD electron spectra with much higher resolution than is possible by the VMI technique [red and black lines in Fig.~2 b)]. Clearly, the peak structure is more complex than that for ICD after resonant optical excitation ($h\nu=21.6$~eV, cyan line). There, electrons are mostly produced by decay of pairs of He$^*$'s in the $^1\Sigma_g$ state correlating to two atoms in 1s2s\,$^1$S singlet states~\cite{LaForge:2021}. In contrast, both singlet and triplet states of He $e$-He can be excitated by electron impact. A detailed peak analysis of the new ICD spectra (SM Fig.~3) shows that they mainly contain 7 peaks, where the dominant ones [labeled as A, B, C in Fig.~2 b)] are related to He$^*$'s in the metastable 1s2s\,$^3$S state. The smaller peaks D, E, F only appear at $h\nu >46.6$~eV (SM Fig.~3 c) and are attributed to ICD involving higher-lying states. We ascribe the dominant role of the 1s2s\,$^3$S state to its long lifetime $\sim15~\mu$s~\cite{McKinsey:2003}, whereas all higher excited states rapidly decay by droplet-induced relaxation~\cite{Mudrich:2014,Asmussen:2021} and fluorescence emission, including the 1s2s\,$^1$S state~\cite{keto:1974,Joppien:1993,Haeften:2001,Haeften:2002,PhysRevA.67.062716}. Furthermore, the 1s2s\,$^3$S and higher triplet states are populated by autoionization and recombination of 1s$n$s-excited states with $n\geq 3$~\cite{Haeften:1997}.

We conclude that this new type of indirect ICD is a slow process involving the following steps: (i) Photoionization followed by photoelectron impact excitation of neutral He, hence formation of He$^*$, (ii) multiple elastic $e$-He scattering and electron-ion recombination, (iii) electronic relaxation of the two He$^*$'s to the lowest excited (mostly triplet) states~\cite{Mudrich:2020,Asmussen:2021}, and (iv) ejection of the He$^*$'s to the droplet surface where they roam until two He$^*$'s encounter and decay by ICD~\cite{Buchenau:1991}. This multistep ICD process is schematically represented in Fig.~3 a). Absorption of two photons by the droplets followed by electron impact excitation and ICD contributes to some extent [Fig.~3 b)].

Additional decay channels are ICD involving He$_2^*$ excimers formed by association of He$^*$'s with neighboring groundstate He atoms and ICD involving He$^*$ in highly excited Rydberg states. These processes appear in the electron spectra of Fig.~2 a) around 10~eV and 19.6~eV, respectively. The latter is only active near the electron-impact excitation threshold ($h\nu=44.5$-$46.5$~eV, see SM Fig.~4), where $e$-He$^+$ recombination occurs at very low electron energy. Direct evidence for ICD where a He$_2^*$ excimer relaxes to the groundstate and a neighboring He$^*$ or He$_2^*$ is ionized is obtained from electron and ion VMI's recorded in coincidence with ions and electrons, respectively [see SM Fig.~5].

Our assertion of the essential role of the new one-photon-induced indirect ICD process is further confirmed by measurements of the ICD electron yield as a function of the intensity of radiation, see SM Fig.~7. At $h\nu=46.5$~eV we find a power-law dependence with exponent $\alpha=1.0$ indicating single-photon absorption, whereas for resonant two-photon excitation per droplet at $h\nu$=$21.0$~eV we find $\alpha=1.3$. At $h\nu=50$~eV, $\alpha=1.15$ indicating that both one- and two-photon absorption contributes to the ICD signal.

\begin{figure}[!h]
\center
\includegraphics[width=0.8\columnwidth]{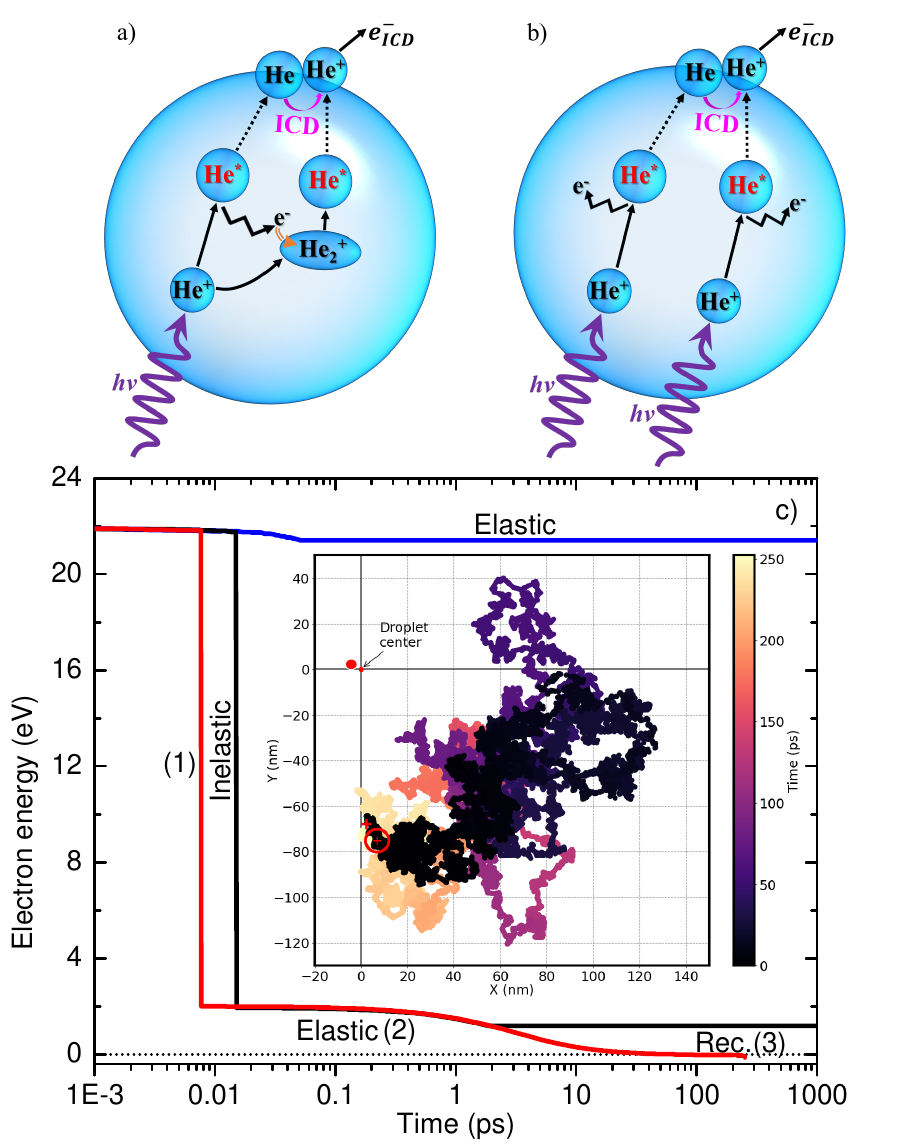}\caption{\label{fig3} a) Illustration of the main ICD mechanism in He droplets induced by photoelectron impact excitation followed by electron-He$^+$ recombination. b) ICD induced by two-photon absorption in a He droplet followed by two electron-impact excitations. c) Electron trajectory simulation for a He droplet of radius $R=150$~nm at $h\nu=46.5$~eV. The inset shows the real-space trajectory projected onto the $xy$-plane; the red circle indicates the (frozen) position of the photoion from where the electron is emitted and which it recombines with at the end of the trajectory.}
\end{figure}

To further establish the multistep mechanism leading to indirect ICD, we performed 3D classical Monte-Carlo trajectory simulations (MCT) of electron propagation inside the He nanodroplets. Binary elastic and inelastic collisions are taken into account according to the known differential cross sections~\cite{Register:1980,Ralchenko:2008}. For details see SM Sec.~IV and V. Fig.~3~c) shows the evolution of the total electron energy along three selected trajectories. When the electron undergoes only elastic scattering, it rapidly reaches the droplet surface and escapes with insignificant loss of energy (blue line). However, given the atomic density of He droplets close to that of bulk superfluid He (0.022~\AA$^{-3}$), the electron has a high probability of undergoing an $e$-He inelastic collision after about 10~fs, thereby suddenly losing $\geq20.6$~eV of its kinetic energy [black and red lines, step (1)]. Subsequently, the electron undergoes multiple elastic collisions with He atoms (2) which leads to a slow, friction-like damping within about 10~ps. Despite the large mismatch of electron mass and He atomic mass (1/7300), the resulting diffusion-like electron motion can be fully stopped given the large sizes of He droplets $R>20$~nm we consider here. In this case it is eventually drawn back to its parent He$^+$ ion via Coulomb attraction and recombines with it (3) after about 100~ps (red line). Note that these values are in good agreement with previous findings for bulk liquid He~\cite{McKinsey:2003}. The inset shows the projection of the trajectory leading to recombination onto the $xy$-plane. Time after ionization is encoded in the line color. Trajectories where two He$^*$'s are formed in one droplet by electron impact and by recombination to the 1s2s\,$^3$S state are counted as ICD events.

\begin{figure}[!h]
\center
\includegraphics[width=0.8\columnwidth]{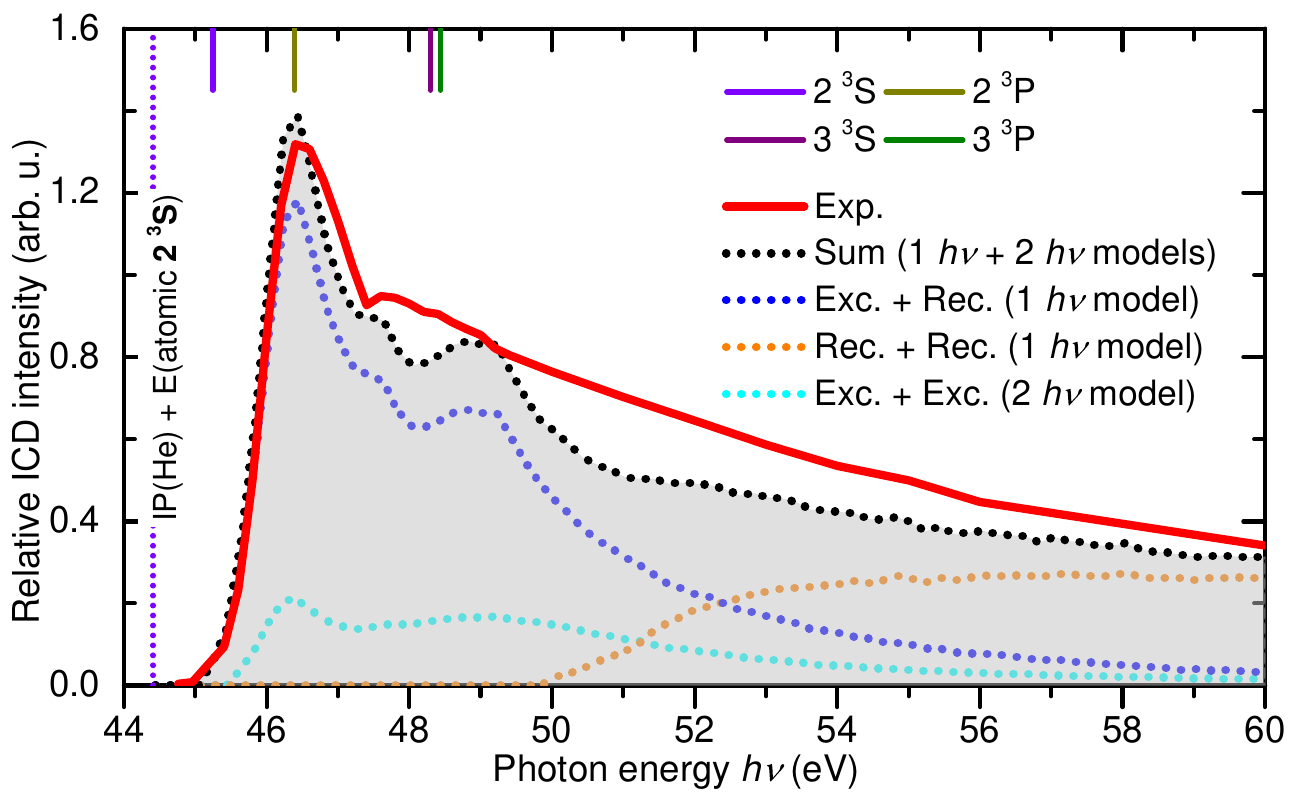}\caption{\label{fig4} Simulated and experimental yields of ICD electrons in proportion to photoelectrons as a function of $h\nu$ for He droplets of radius $R=50$~nm. Colored dashed lines indicate different channels leading to ICD in the simulation. Colored vertical solid lines indicate He electron impact excitation energies shifted by 0.84~eV. The vertical purple dashed line indicates the He electron impact excitation threshold at $h\nu = 44.4$~eV.}
\end{figure}

The simulated numbers of ICD events following one- and two-photon ionization in proportion to the number of directly emitted electrons are shown in Fig.~4 as blue, orange and cyan dashed lines, respectively, for a He droplet of size $R=50$~nm and for $h\nu$'s tuned across the He impact excitation threshold (see details in SM Sec.~IV). They are summed (black dashed lines in Fig.~4) and compared to the experimental ICD rate normalized to the rate of detected photoelectrons (red solid line). To reproduce the sharp ICD peak around $h\nu = 46.5$~eV, He electron impact excitation energies are shifted by 0.84~eV with respect to atomic-level energies. This shift is consistent with blue-shifts of excited states observed in optical and $e$-He impact excitation spectra~\cite{Joppien:1993,Shcherbinin:2019}.
The best fit of the simulated and experimental data is obtained when scaling down the channels leading to ICD, electron impact excitation followed by $e$-He$^+$ recombination (blue dashed lines in Fig.~4) and electron impact ionization followed by recombination of the two emerging electrons (orange dashed lines in Fig.~4) by factors 0.3 and 0.1, respectively (see SM Sec.~IV).

Overall, the simulation confirms our interpretation that ICD is mainly initiated by one-photon ionization. At $h\nu=45$-$53$~eV, impact excitation of He by the photoelectron and $e$-He$^+$ recombination is the main pathway to ICD. At $h\nu$ $\gtrsim$ 53~eV, the electron impact-ionization cross section exceeds those for electron impact-excitation and recombination of two electrons dominates. At $h\nu>65.4$~eV (not shown), multiple electron impact excitation and ionization can generate two He$^*$'s in a droplet thereby further enhancing ICD. Two-photon absorption causing the photoemission of two electrons which both impact excite He$^*$'s in the same droplet only plays a minor role in this droplet size range 5-75~nm but may become important for larger droplets and higher radiation intensities~\cite{Ovcharenko:2014,Ovcharenko:2020}. 

We point out that this type of ICD is not restricted to pure He droplets; in He droplets doped with lithium (Li) atoms, ICD also occurs between a He$^*$ and a Li atom thereby creating a Li$^+$ ion and an ICD electron~\cite{Ltaief:2019}. The yield of Li$^+$ ions produced in this way at $h\nu>44.4$~eV follows essentially the same $h\nu$ dependence [see SM Fig.~8], highlighting the relevance of this ICD mechanism also for heterogeneous systems.

\begin{figure}[!h]
	\center
	\includegraphics[width=0.8\columnwidth]{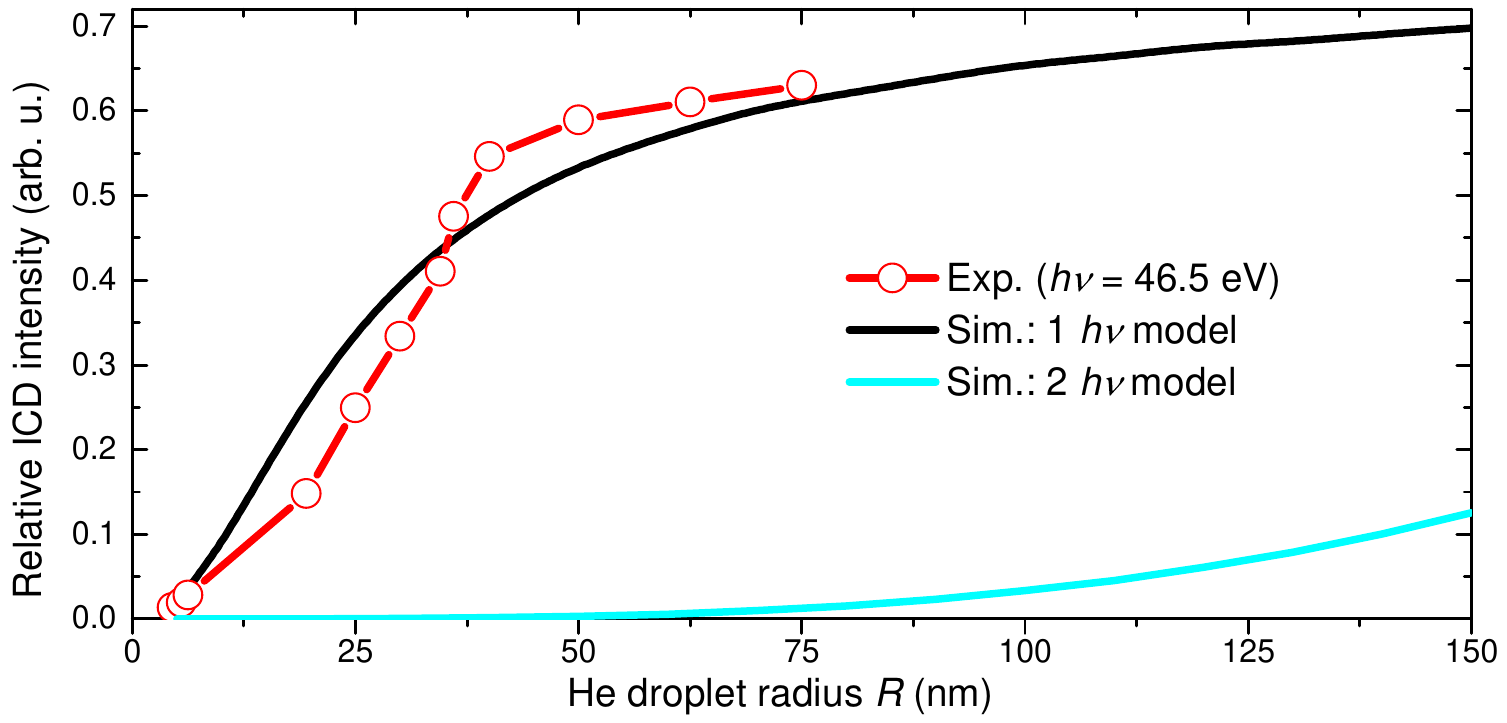}\caption{\label{fig5} Experimental vs. simulated yields of ICD electrons normalized to all electrons as a function of He droplet radius $R$ at $h\nu = 46.5$~eV.}
\end{figure}

In conclusion, in large He droplets irradiated by XUV light at $h\nu\gtrsim44.4$~eV, ICD of pairs of He$^*$'s formed by electron impact and by electron-ion recombination is an important relaxation mechanism generating electrons with characteristic energy $\approx 15$~eV. High-resolution electron spectra reveal that mainly the lowest metastable excited state of He contributes, indicating that higher-lying states relax radiatively and non-radiatively prior to ICD. This sets the timescale for this type of ICD to nanoseconds or even longer.
For droplet radii $R\gtrsim 40$~nm, ICD even becomes the dominant electron emission channel. Fig.~5 shows the experimental and simulated (see SM Sec. IV) yields of ICD electrons normalized to the sum of all detected electrons as a function of $R$. Starting at $R\approx5$~nm, both experimental and simulated ICD electron yields monotonously rise and reach a value of about 0.65 at $R=75$~nm. We refrain from analyzing larger droplets as more complex processes will contribute such as multi-photon absorption (cyan curve in Fig.~5), shadowing, nanofocusing~\cite{ban2020photoemission}, and trapping of electrons at the droplet surface~\cite{Farnik:1998}. Future studies should aim at studying the dynamics of this indirect ICD~\cite{LaForge:2021,Ziemkiewicz:2015} and quantifying the role of competing fluorescence decay channels~\cite{Haeften:2023}. Comparative studies of other condensed-phase systems should be done to confirm the general relevance of this type of ICD for generating slow electrons, which can cause radiation damage in biological matter~\cite{Alizadeh:2015}.

\section{Acknowledgements}
M.M. and L.B.L. acknowledge financial support by Deutsche Forschungsgemeinschaft (project BE 6788/1-1), by the Danish Council for Independent Research Fund (DFF) via Grant No. 1026-00299B and by the Carlsberg Foundation. TF acknowledges support by the Deutsche Forschungsgemeinschaft (DFG, German Research Foundation) via SFB 1477 ``light–matter interactions at interfaces'' (project number 441234705) and via the Heisenberg program (project number 436382461). SRK thanks Dept. of Science and Technology, Govt. of India, for support through the DST-DAAD scheme and Science and Eng. Research Board. SRK, KS and SD acknowledge the support of the Scheme for Promotion of Academic Research Collaboration, Min. of Edu., Govt. of India, and the Institute of Excellence programme at IIT-Madras via the Quantum Center for Diamond and Emergent Materials. SRK gratefully acknowledges support of the Max Planck Society's Partner group programme. M.M. and S.R.K. gratefully acknowledge funding from the SPARC Programme, MHRD, India. The research leading to this result has been supported by the project CALIPSOplus under grant agreement 730872 from the EU Framework Programme for Research and Innovation HORIZON 2020 and by the COST Action CA21101 ``Confined Molecular Systems: From a New Generation of Materials to the Stars (COSY)''.

%

\end{document}


\title{Supplementary Material\linebreak
Efficient Indirect Interatomic Coulombic decay induced by photoelectron impact excitation in large pure He nanodroplets}

\author{L. Ben Ltaief \textit{et al.}}

\maketitle

\renewcommand\refname{SM References}
\renewcommand{\figurename}{SM Fig.}

\date{}

 \section{Electron spectra}
In this section we present additional electron spectra recorded at various He droplet sizes and photon energies to support the discussion given in the main text.  

\subsection{Electron energy-loss spectra}
\begin{figure}[!h]\center
\includegraphics[width=0.44\columnwidth]{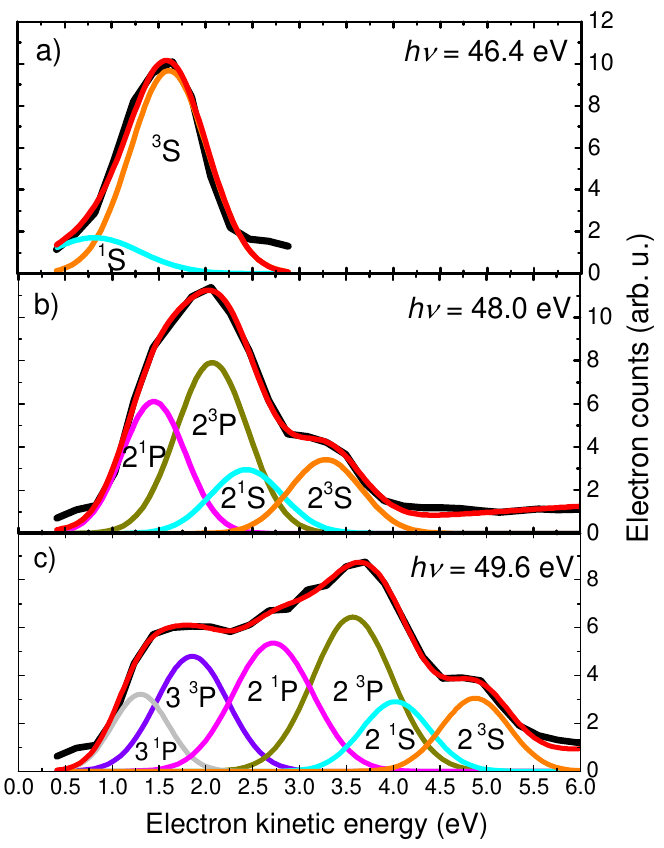}\caption{\label{fig1} Electron energy-loss spectra measured by VMI for He droplets of radius $R=75$~nm and at various photon energies $h\nu$. Colored lines are the individual components of multi-gaussian peak fits.}
\end{figure}
From the ICD electron spectra recorded with high resolution (Fig.~2 b) in the main text) we infer which excited states in He droplets decay by ICD. In contrast to ICD induced by resonant optical excitation~\cite{Ovcharenko:2020}, predominantly the metastable 1s2s\,$^3$S state contributes to ICD although other states are expected to be excited by electron impact based on the known impact-excitation cross sections for He atoms~\cite{Ralchenko:2008}. Naturally the question arises, which initial states with excitation energy $E_*$ are impact excited in He droplets. To answer this question, we analyze the electron energy-loss (EEL) feature in the electron spectra showing up at electron energies $h\nu - E_i - E_*$, where $E_i=24.6$~eV is the ionization energy of He. SM Fig.~1 shows the electron energy-loss spectra (EELS) for He droplets of radius $R\approx75$~nm. The labelled gaussian peak fits illustrate the contribution of individual excited states. Note that these measurements were done using the VMI spectrometer. Therefore the energy resolution is limited and the systematic uncertainly is larger than for the spectra measured with the electron analyzer (Fig.~2 b) in the main text).

At $h\nu=46.4$~eV [SM Fig.~1 a)], we indeed find that predominantly the 1s2s\,$^3$S state is excited. Note, that the structure of the EELS changes as a function of droplet size, see the electron spectra shown in Fig.~2 a) of the main text in the range 0.2-2.5~eV; The two distinct peaks due to impact excitation of the 1s2s\,$^3$S and 1s2s\,$^1$S states of He observed for small He droplets merge into one feature for $R\gtrsim20$~nm. It appears that in large droplets the probabilities of exciting various droplet states are altered with respect to the atomic impact-excitation cross sections~\cite{Ralchenko:2008}. Note that optical excitation spectra of He droplets have previously also been found to qualitatively change structure when the size of the droplets is varied from $N<10^6$ ($R<22$~nm) to $N>10^7$ ($R>48$~nm) He atoms per droplet~\cite{Haeften:2011,Haeften:2023}. For higher photon energies $h\nu\geq48$~eV [SM Fig.~1 b) and c)], also singlet states are excited with similar probabilities. However, their contribution to the ICD electron spectra is small at all photon energies, indicating that singlet states partially decay prior to ICD. Overall, the EELS are in decent agreement with previous measurements done for small He droplets~\cite{Shcherbinin:2019}.

\subsection{ICD electron spectra at higher photon energies}
\begin{figure}[!h]
	\center
 \includegraphics[width=0.5\columnwidth]{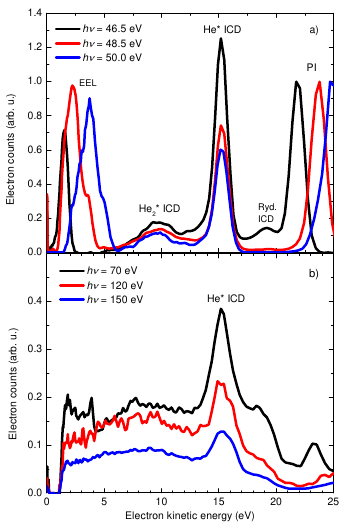}\caption{\label{fig2_SM} Electron spectra of He droplets of radius $R=50$~nm measured by VMI at different photon energies. The spectra shown in the upper panel are normalized to the He photoline; the spectra shown in the lower panel are normalized to the photon flux.}
\end{figure}
Additional electron spectra of He droplets of radius $R=50$~nm measured at various photon energies using the VMI spectrometer are shown in SM Fig.~2. The spectra recorded at few eV above the threshold for electron-impact excitation, shown in SM Fig.~2 a), display the same features as that shown in Fig.~2 a) in the main text. The positions of ICD peaks around 10, 15, 19.6~eV are independent of photon energy. At higher photon energies $h\nu=70$, $120$, and $150$~eV [SM Fig.~2 b)], the ICD peak at 15~eV remains the dominant feature which highlights the importance of the ICD process discussed in this work. At $h\nu=70$, additional features are present in the electron spectrum due to simultaneous ionization and excitation of a He atom forming He$^{+*}$, which decays by another type of ICD~\cite{Shcherbinin:2017,Kazandjian:2018}. The signal minimum in the range 0-1~eV is characteristic for electron emission out of large He droplets and has been discussed as the manifestation of a conduction band for electrons forming in large He droplets, also known as the energy barrier to injection of free electrons into bulk liquid He~\cite{Wang:2008,mauracher2018cold,Shcherbinin:2018}.

\subsection{High-resolution ICD electron spectra}
 \begin{figure}[!h]
	\center
	\includegraphics[width=0.8\columnwidth]{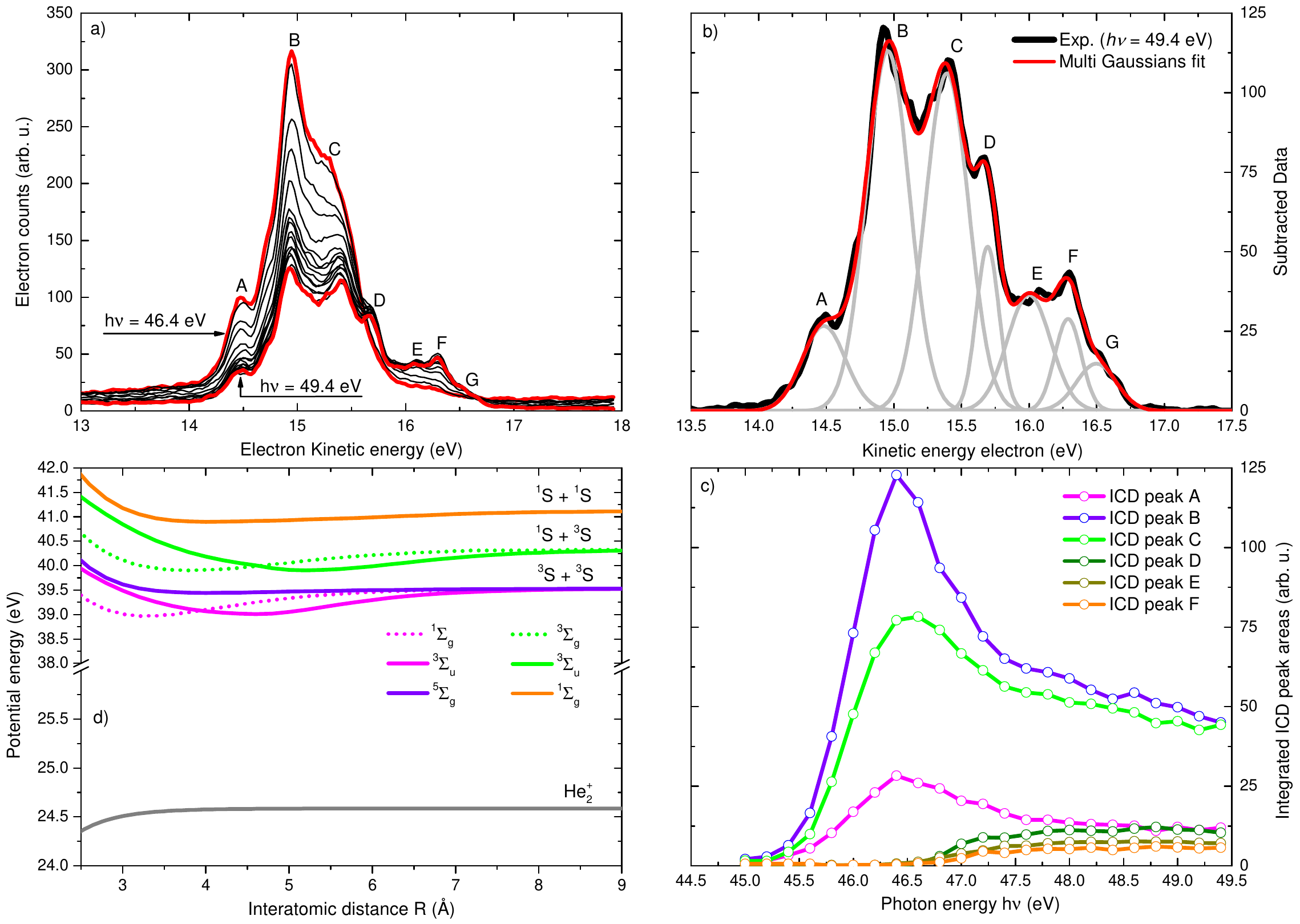}\caption{\label{fig3} a) High-resolution electron spectra measured for He nanodroplets of radius $R=50$~nm around the ICD feature in the kinetic energy range $13.0-18.0~eV$. The photon energy was varied in the narrow range $h\nu=46.4$-$49.4$~eV in steps of 0.2~eV. b) ICD electron spectrum measured at $h\nu=49.4$~eV including a multi-gaussian peak fit consisting of the sum of 7 gaussians. c) Photon-energy dependence of peak integrals of the most well-resolved ICD peaks seen in a) and b). d) Potential energy curves for pairs of He$^{*}$ atoms in their lowest excited states 1s2s\,$^3$S and 1s2s\,$^1$S~\cite{LaForge:2021}. }
\end{figure}
 SM Fig.~3 a) shows a series of high-resolution electron spectra measured for large He nanodroplets of $R=50$~nm at variable photon energy in the range  $h\nu=46.4$-$49.4$~eV. These spectra were recorded using the hemispherical electron analyzer. The overall ICD signal intensity drops when tuning $h\nu$ from $46.4$ to $49.4$~eV. Up to 7 ICD peaks A-G are discernible. A, B, C peaks are the most dominant, whereas the D, E, F smaller peaks only appear at $h\nu >46.6$~eV, see SM Fig.~3 c). The ICD feature at $h\nu=49.4$~eV is well reproduced by a multi-gaussian fit comprising 7 gaussians, see SM Fig.~3 b). The positions and the assignment of the peaks are given in Table 1 ($E_\mathrm{fit}$). For comparison, Table 1 contains the electron energy values expected for the ICD process for the unshifted atomic and molecular excited states ($E_\mathrm{lit.}$). The fit results agree with the calculated ICD electron energies within the resolution of the spectrometer ($\leq$ 0.1~eV) for those lines where He$^*$-He$^*$ pair potentials are available in the literature~\cite{LaForge:2021}, see SM Fig.~3 d). The assignment of peaks D and E is tentative and G remains unassigned. 
 
\begin{table}[h!]
\centering
\caption{Assignment of ICD peaks labeled in Fig 3 and their peak positions $E_\mathrm{fit}$ in comparison with the calculated values $E_\mathrm{lit}$ based on literature atomic term values and the potential energy curves for pairs of He$^{*}$
atoms in their lowest excited states 1s2s\,$^3$S and 1s2s\,$^1$S states~\cite{LaForge:2021}.}

\begin{tabular}{c c c c c c c} 
 \hline
Peaks& Assignment & $E_\mathrm{lit.}$ & $E_\mathrm{fit}$\\[0.1ex] 
 \hline \\
 A &$^{3,1}\Sigma_{u,g}$($^{3}S-$$^{3}S$)& 14.43&14.48\\\\ 
 B &$^{5}\Sigma_{g}$($^{3}S-$$^{3}S$)& 14.87&14.95\\
   & $^{3}S-$$^{3}S$& 15.05&14.95\\\\
 C &$^{3}\Sigma_{u,g}$($^{3}S-$$^{1}S$)&15.33 &15.39\\
   &$^{3}S-$$^{1}S$&15.85&15.39\\\\
 D & - & - &15.69\\
  &$^{3}S-$$^{3}P$&16.20&15.69\\\\
 E & - & - &15.99\\
  &$^{3}S-$$^{1}P$&16.45&\\\\ 
 F&$^{1}\Sigma_{g}$($^{1}S-$$^{1}S$)&16.32&16.28\\
  &$^{1}S-$$^{1}S$&16.64&16.28\\
 G& - & - &16.50\\
 [0.5ex] 
 \hline
\end{tabular}
\label{table:1}
\end{table}

\subsection{ICD involving He atom in Rydberg states}
\begin{figure}[!h]
	\center	\includegraphics[width=0.5\columnwidth]{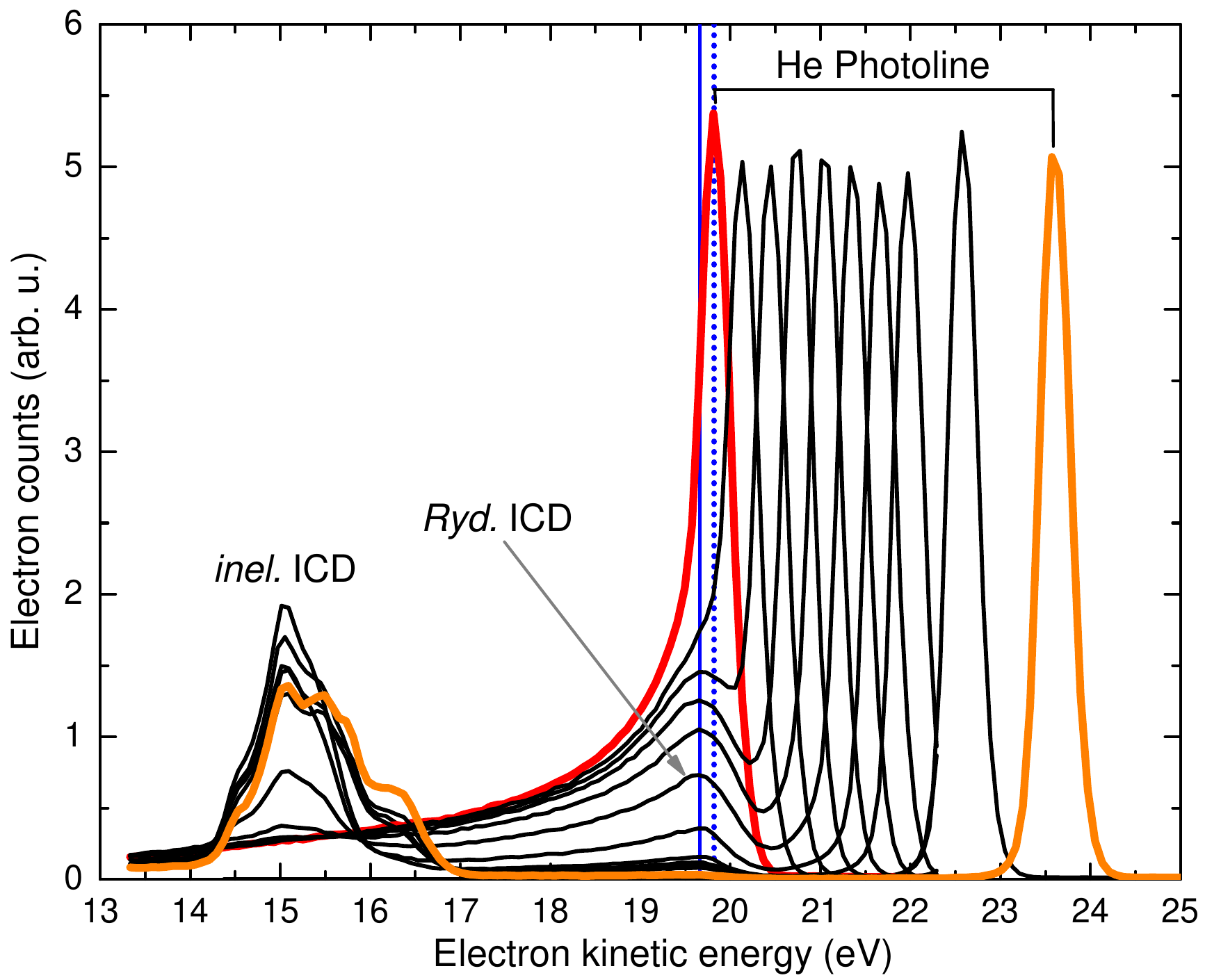}\caption{\label{fig4_SM} High-resolution electron spectra measured for He droplets of radius $R=50$~nm and in the narrow photon energy range $h\nu=44.3$-$49.0$~eV. From $h\nu=44.3$ to $46.7$~eV, $h\nu$ is incremented by 0.3~eV. The solid vertical blue line indicates the position of the local maximum at 19.6~eV. The dotted vertical blue line indicates the expected electron energy of an electron emitted by ICD involving an excited He$^*$ atom in a high-lying Rydberg state near $E_i$ and a He$^*$ in the 1s2s\,$^3$S state.}
\end{figure}
The feature observed around 19.6~eV in the electron spectra of Fig.~2 a) of the manuscript appears for droplets with radius $R\geq20$~nm, and becomes more and more pronounced for larger droplets ($R=75$~nm). It is attributed to ICD that occurs between one He$^{*}$ in the lowest excited state 1s2s\,$^3$S and one He$^{*}$ in a highly excited Rydberg state near the ionization threshold. This `Rydberg ICD' feature is only present near the electron-impact excitation threshold ($h\nu=44.5$-$46.5$~eV) where $e$-He$^+$ recombination occurs at very low electron energy as it can be seen in SM Fig.~4 showing electron spectra measured by the hemispherical electron analyser for He droplets of radius $R=50$~nm in the narrow photon-energy range $h\nu=44.3$-$49.0$~eV. Presumably, at low energy the photoelectron promptly recombines with the parent He$^+$ ion without diffusing through the droplet and before formation of He$_2^+$ sets in. 
 
\subsection{ICD involving He$_2^*$ excimers}
\begin{figure}[!h]
	\center
	\includegraphics[width=0.7\columnwidth]{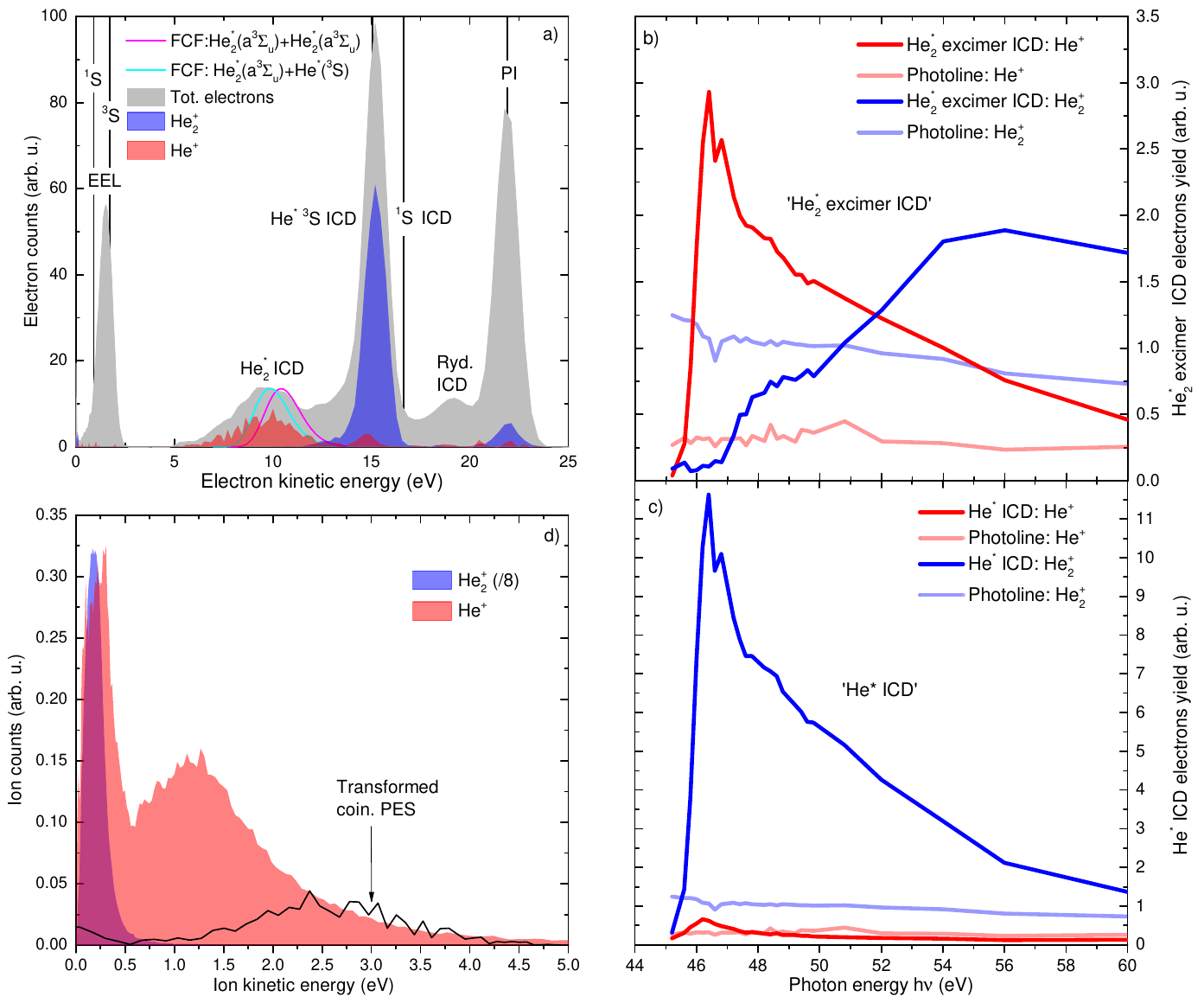}\caption{\label{fig5} a) Electron spectra measured in coincidence with He$^+$ (red area) and He$_2^+$ (blue area) at $h\nu=46.5$~eV for He droplets with radius $R=50$~nm. The gray area shows the total electron spectrum. The cyan and pink lines indicate Franck-Condon (FC) profiles simulated for the lowest triplet excited state a$^{3}\Sigma_{u}$ of He$_2^*$ interacting with a He atom in the 1s2s\,$^3$S state and with the a He$_2^*$ excimer in the $a\,^{3}\Sigma_{u}$ state, respectively. b) Photon energy-dependent yield of electrons measured in coincidence with He$^+$ (red curve) and He$_2^+$ (blue curve), and which are leading to the He$_2^*$ excimer ICD feature seen around 10~eV in the electron spectra of Fig.~5 a). c) Photon energy-dependent yield of electrons measured in coincidence with He$^+$ (red curve) and He$_2^+$ (blue curve), and which are leading to the He$^*$ ICD feature seen around 15~eV in the electron spectra of Fig.~5 a). The light red and blue curves in Fig.~5 b) and c) are yield of photoelectrons detected in coincidence with He$^+$ and He$_2^+$, respectively. All the $h\nu$-dependent ICD signal shown in Fig.~5 b) and c) are normalized to the flux of the photon beam. d) He$^+$ (red spectrum) and He$_2^+$ (blue spectrum) ion kinetic energy distributions measured at $h\nu$=46.5~eV and for He droplets with droplet radius of 50~nm. The curves plotted in black is a linear transformation of the electron spectrum measured in coincidence with He$^+$ (red spectrum in Fig.~5 a)).}
\end{figure}
Next to the main sharp ICD peak at 15~eV another broad ICD feature is present in electron spectra centered around 10~eV, see Fig.~2~a) in the main text. These electrons are formed by ICD involving He$_2^*$ excimers whose excitation energy is lower than that of the He$^*$ atom by about 5~eV. SM Fig.~5~a) shows electron spectra for He droplets of radius $R=50$~nm measured at $h\nu=46.5$~eV by recording all emitted electrons (grey area) and electrons in coincidence with He$^+$ (red area) and He$_2^+$ (blue area). Counter-intuitively (at first glance), the majority of electrons emitted with kinetic energy 15~eV -- by ICD involving two He$^*$'s -- are detected in coincidence with He$_2^+$ ions, whereas those electrons formed by ICD involving He$_2^*$ excimers (feature around 10~eV) are mainly detected in coincidence with He$^+$. This is further illustrated in Fig.~5 b) and c) showing the photon-energy dependent peak integrals of the excimer ICD peak and the atomic ICD peak, respectively, for $e$-He$^+$ coincidences (red lines) and for $e$-He$_2^+$ coincidences (blue lines). The photon energy is varied in the range $h\nu=44.0$-$60.0$~eV. 
ICD involving He$_2^*$ excimers can take place between one excimer He$_2^*$ state and one excited atom He$^*$ or between two excimers He$_2^*$ in the reactions
\begin{eqnarray}
\mathrm{He}_\mathrm{2}^\mathrm{*} + \mathrm{He}^\mathrm{*}&\xrightarrow{}& \mathrm{He}_\mathrm{2} + \mathrm{He}^\mathrm{+}  + \mathrm{e}_\mathrm{ICD}^\mathrm{-} ,\\
\mathrm{He}_\mathrm{2}^\mathrm{*} + \mathrm{He}_\mathrm{2}^\mathrm{*} &\xrightarrow{}& \mathrm{He}_\mathrm{2} + \mathrm{He}_\mathrm{2}^\mathrm{+} + \mathrm{e}_\mathrm{ICD}^\mathrm{-} .
\end{eqnarray}
To confirm the interpretation of the broad feature around 10~eV, we numerically calculate Franck-Condon profiles for the transition from the lowest triplet state $a^3\Sigma_{u}$ of He$_2^*$ to the He$_2$ ground state leading to ionization of He$^*$ in the 1s2s\,$^3$S state or to ionization of He$_2^*$ in the $a^3\Sigma_{u}$ state. This was done using the computer program BCONT~\cite{bcont} and the potential energy curves given in Ref.~\cite{Fiedler:2014}. The resulting Franck-Condon profiles are shown in SM Fig.~5~a) as cyan and magenta solid lines. Both profiles overlap with the experimental `excimer ICD' feature, indicating that both processes probably contribute.

Further evidence for ICD according to the reactions (2) and (3) is obtained from the ion kinetic-energy distributions measured by electron-ion coincidence VMI. SM Fig.~5 d) shows spectra of He$^+$ (red area) and He$_2^+$ (blue area) recorded in coincidence with electrons. Both ion spectra exhibit a sharp peak in the kinetic-energy range 0-0.7~eV. Additionally, the He$^+$ ion spectrum features a broad maximum around 1.2~eV reaching out to 5~eV. This feature can be related to the `excimer ICD' feature in the electron spectra [broad maximum around 10~eV in Fig.~5 a)]. It mainly reflects the He$_2^*$ vibrational wave function which transforms into the He$^+$ ICD ion kinetic energy distribution by reflection at the repulsive hard-core potential of the He$_2$ ground state upon ICD. One of the dissociating He atoms subsequently collides with the adjacent He$^+$ ion, similarly to another recently studied ICD process~\cite{Shcherbinin:2017,Kazandjian:2018,Wiegandt:2019}. This process is schematically shown in SM Fig.~6. The black line in SM Fig.~5 d) is obtained by linear transformation of the electron spectrum measured in coincidence with He$^+$ [red area in SM Fig.~5 a)] according to $E_i=(E_\mathrm{He^*-ICD}-E_e)/2$.
Here, $E_e$ is the electron energy and $E_\mathrm{He^*-ICD}=15$~eV is the ICD electron energy for reaction (1). The factor $1/2$ accounts for equal sharing of the kinetic energy released to the two dissociating He atoms. The high-energy edge of this distribution matches the one of the He$^+$ ion kinetic-energy spectrum quite well indicating that part of the He$^+$ ions are indeed ejected from the droplets by a direct head-on collision of one of the dissociating He atoms with the ICD He$^+$ ion without undergoing further scattering. In He-He$^+$ collisions at non-zero collision parameter the kinetic energy of the He is only partly transferred to the He$^+$ which explains the large contribution of lower energies in the measured He$^+$ energy distribution. Furthermore, after the collision the He$^+$ ion can further scatter at surrounding He atoms in the droplet and thus lose more energy. This might explain the narrow distribution peaked at 0.3 eV. More refined simulations would be needed to come up with a full quantitative description of this system. He$_2^+$ ions are ejected from He droplets with a kinetic energy around 0.3~eV [blue area in SM Fig.~5 d)] presumably by a non-thermal process driven by vibrational relaxation of He$_2^+$ initially formed in vibrationally excited states ~\cite{callicoatt1998fragmentation,Shcherbinin:2017}. 

\begin{figure}[!h]
	\center	\includegraphics[width=0.5\columnwidth]{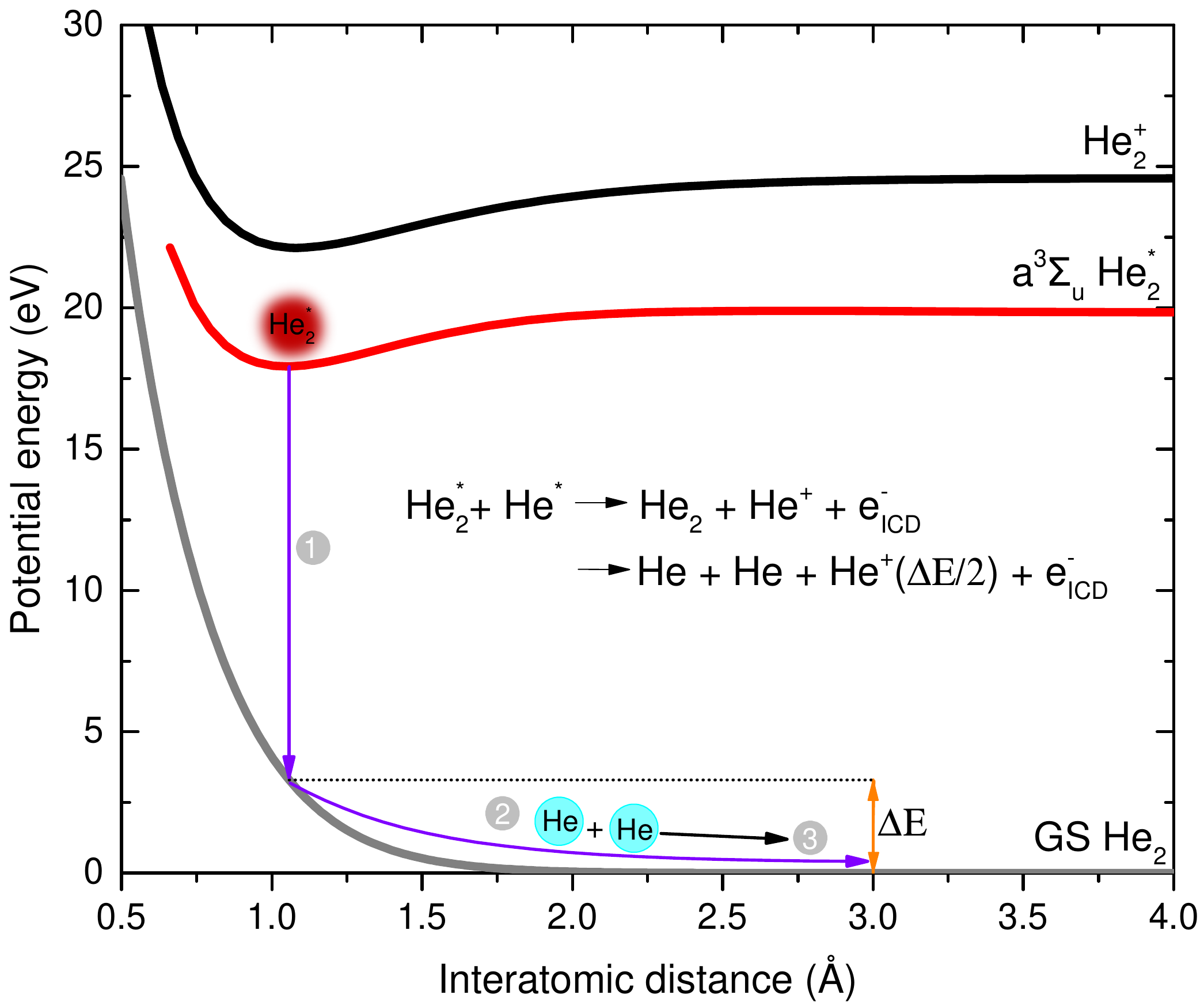}\caption{\label{SM_Fig6} Potential energy curves of ground state He$_2$ (grey line), He$_2^*$ excimer in the lowest $a\,^3\Sigma_u$ state (red line) and of He$_2^+$ (black line). When a He$_2^*$ formed in a He droplet decays by ICD it relaxes to the He$_2$ ground state (GS, step 1). In turn, the He$_2$ dissociates into two energetic neutral He atoms (step 2) where one of them subsequently undergoes a collision with the adjacent He$^+$ ICD ion (step 3) and transfer its kinetic energy $\Delta E/2$ to the He$^+$ ion formed in step 1.}
\end{figure}

\section{Dependence of the ICD signal on photon beam intensity}
\begin{figure}[!h]
	\center	\includegraphics[width=0.5\columnwidth]{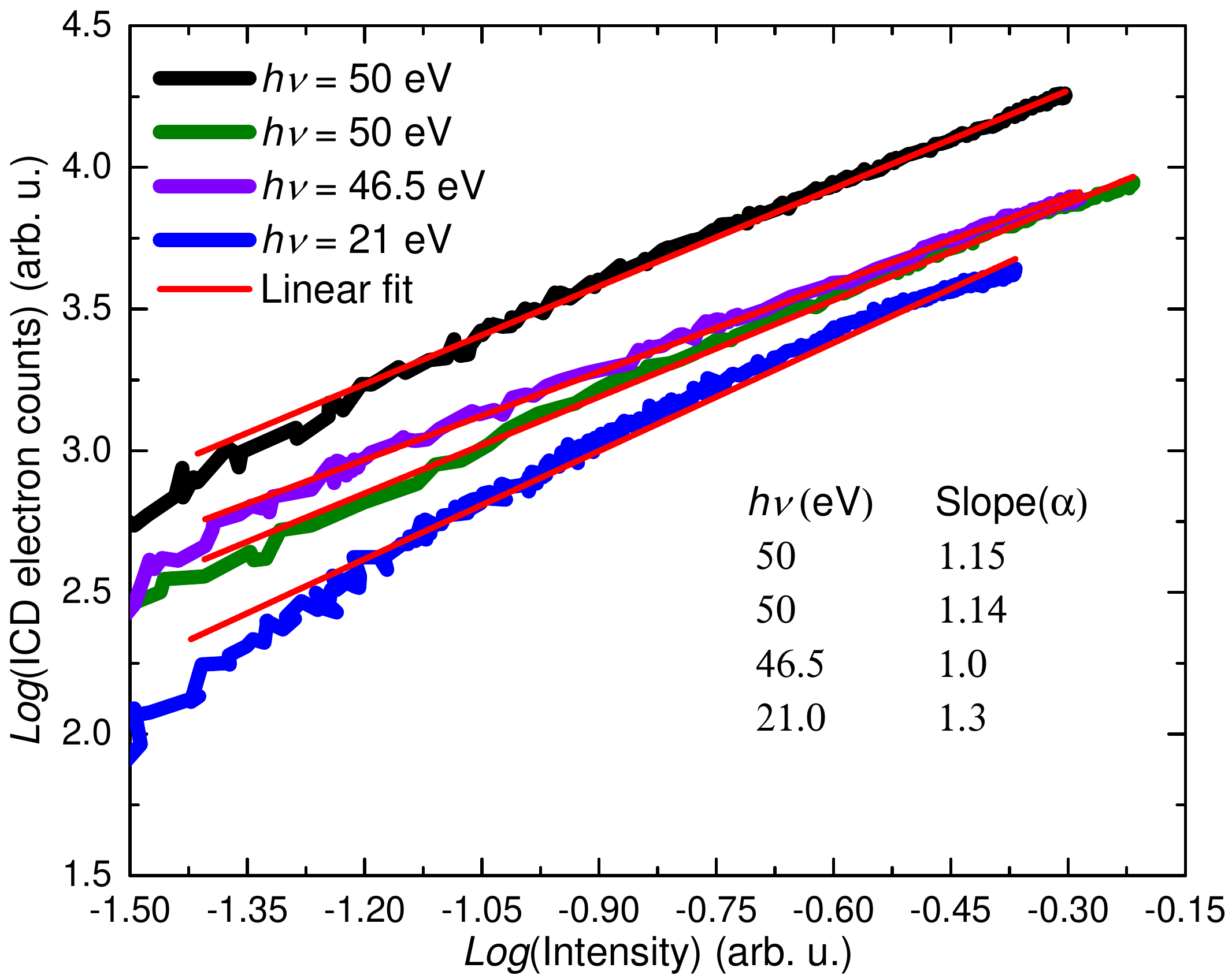}\caption{\label{fig7} Yield of ICD electrons measured for He droplets with radius ($R=75$~nm) and for different photon energies $h\nu=21.0$~eV (blue line), $h\nu=46.5$~eV (purple line) and $h\nu=50$~eV (green and black lines). The photon beam intensity was varied by gradually closing the exit slit of the monochromator of the beamline. The black and green signals were measured at the same droplet beam conditions and photon energy but on different days.}
\end{figure}
An intriguing question is whether different ICD mechanisms (one-photon ionization followed by $e$-He$^+$ recombination, two-photon absorption by the same droplet followed by impact excitation, multi-photon processes) can be discerned by the dependence of their yield $Y\propto I^\alpha$ on the photon beam intensity $I$. In Ref.~\cite{Ovcharenko:2014} a linear dependence on photon flux was found for ICD induced by multiple excitation of He droplets by intense FEL pulses. However, in the present experiment the intensity is several orders of magnitude lower but multiple photon absorption can still happen owing to the large size of the droplets. Fig.~7 shows photon-beam intensity-dependent yields of ICD electrons measured using the hemispherical analyzer for three different photon energies. At $h\nu=21.0$~eV, He droplets are resonantly photoexcited and ICD proceeds when two He$^*$'s or more are created in one droplet. At $h\nu=46.5$~eV, ICD is presumably induced by one-photon photoionization following by inelastic scattering and $e$-He$^+$ recombination. At $h\nu=50$~eV we reckon that the latter process and additionally two-photon ionization and inelastic scattering contribute. From linear fits of the signals in log-log representation, we obtain slopes $\alpha=1.0$ at $h\nu=46.5$~eV indicating that ICD is mainly initiated by one-photon resonant excitation, whereas at $h\nu=21.0$~eV we find a slope $\alpha=1.3$ indicating a weakly non-linear dependence. At $h\nu=50.0$~eV, we find $\alpha=1.15$ which we take as an indication that both one- and two-photon absorption contributes to the yield of ICD electrons. Note that these measurements where done by varying the width of a slit installed at the exit of the monochromator. In the range of highest intensity at the highest slit width ($300~\mu$m), we estimate the bandwidth of the transmitted radiation to increase up to $\sim0.070~$eV which might affect the effective absorption rate, in particular at $h\nu=21.0$~eV, where the absorption resonance has a width of $\sim0.2$~eV.

\section{ICD of doped helium nanodroplets}
In addition to the hitherto discussed results, we have also done experiments on ICD in He nanodroplets doped with lithium (Li) atoms and irradiated with photons of energy $h\nu\geq43.5$~eV. Those experiments are conducted at the synchrotron radiation facility ASTRID2, Aarhus. To this end, a crucible containing Li metal is mounted in the He-droplet beam before the spectrometer chamber and heated to $400\,^\circ$C. In previous experiments where He droplets were resonantly photo-excited, we evidenced efficient ICD according to the reaction~\cite{Buchta:2013,Ltaief:2019}
\begin{equation}
\mathrm{He}^\mathrm{*} + \mathrm{Li} \xrightarrow{} \mathrm{He} + \mathrm{Li}^\mathrm{+} + e_\mathrm{ICD}^\mathrm{-}.
\end{equation}
SM Fig.~8 shows the yields of Li$^+$ and He$_2^+$ ions produced by ICD when He nanodroplets of radius $R=75$~nm are ionized near the electron impact excitation threshold in the photon-energy range $h\nu=43.50$-$49.50$~eV. The $h\nu$-dependent yield of Li$^+$ closely follows the yield of He$_2^+$ ICD ions. This shows that ICD induced by photoelectron impact excitation and $e$-He$^+$ recombination not only occurs in pure He nanodroplets but also in He nanodroplets doped with other species.
\begin{figure}[!h]
	\center
\includegraphics[width=0.5\columnwidth]{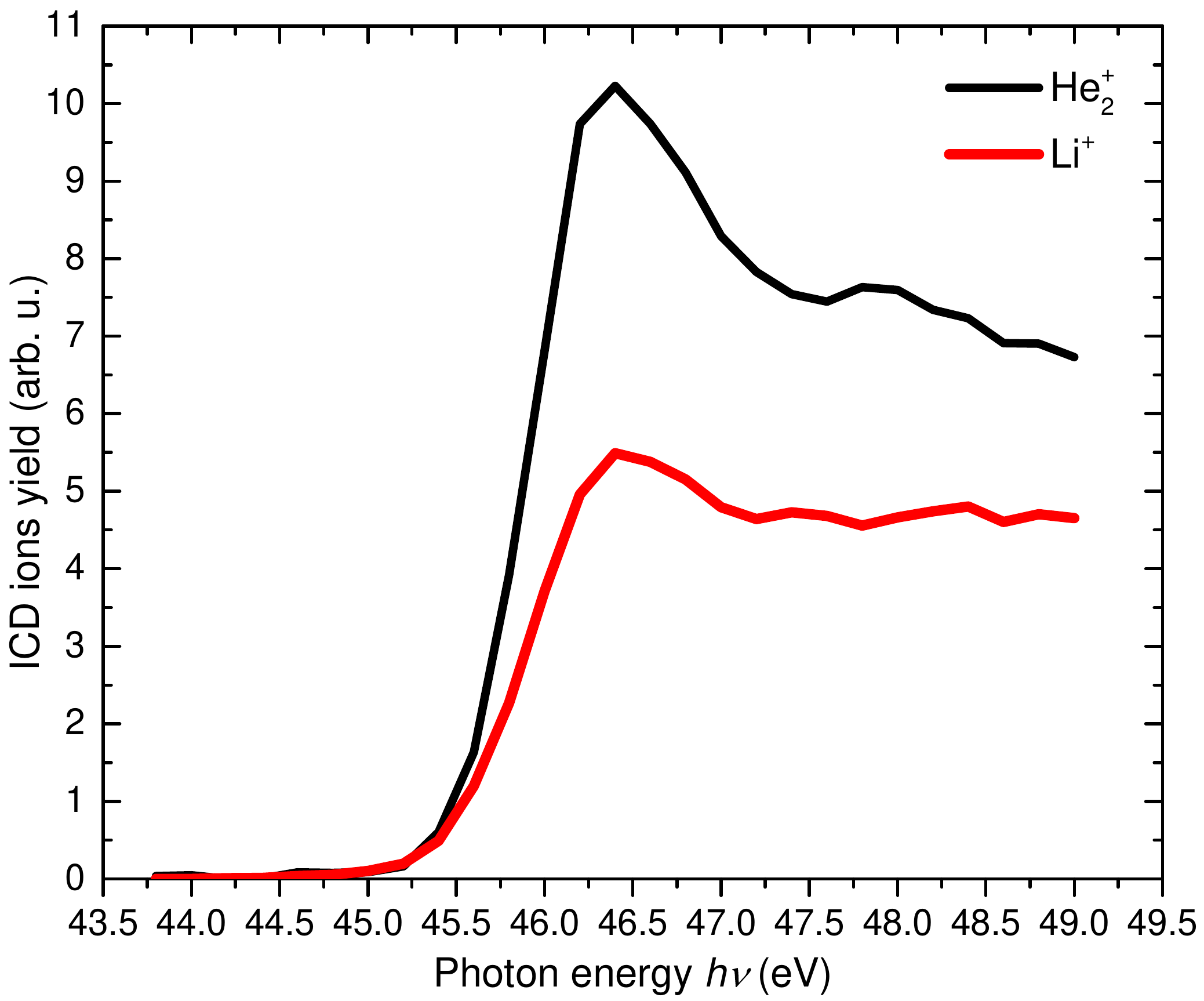}\caption{\label{fig8_SM} Yields of Li$^+$ (red curve) and He$_2^+$ ICD ions (black curve) measured in the photon energy range $h\nu=43.50$-$49.50$~eV for Li-doped He nanodroplets of radius $R=75$~nm. Li$^+$ and He$_2^+$ yields are normalized to the photon flux.}
\end{figure}

\section{Electron trajectory simulation}
The numerical simulations presented in Fig.~3-5 of the main text are carried out as a combination of classical trajectory propagation and Monte-Carlo simulation of the elastic and inelastic $e$-He scattering. Each trajectory is initialized at a random initial position inside a He droplet represented by a sphere of radius $R$ with homogeneous bulk He density. The electron is propagated with an initial kinetic energy $h\nu - E_i$ under the influence of the attractive Coulomb force of the residual ion which is kept fixed at its initial position. The trajectory is terminated when either the electron escapes from the droplet or when the electron returns to the He$^+$ ion. Elastic and inelastic scattering of the electron with He atoms is implemented by the Monte-Carlo method employing energy-dependent inelastic cross sections to the $n=2,\,3\,^{3}$S and $n=2,\,3\,^{3}$P states~\cite{Ralchenko:2008} and energy and angle-dependent elastic scattering cross sections. The double-differential cross sections, determined quantum mechanically via partial wave analysis using the electron-atom potential calculated with standard density-functional theory, are in good agreement with experimental results~\cite{Adibzadeh:2005}. For each value of $R$, $10^5$ trajectories are simulated and analyzed in terms of the final outcome. 

Events of direct emission of an electron with no inelastic collision was counted as photoelectrons. Trajectories where an inelastic collision occurred and the electron returned to the ions were counted as ICD events.
The ICD signal per droplet and following one-photoionization can therefore be expressed as
\begin{equation}
S_\mathrm{ICD(1h\nu)} \approx P_\mathrm{1} \times P_\mathrm{exc} \times P_\mathrm{recomb}
\end{equation}
where $P_\mathbf{1}$ is the probability that a He nanodroplet absorbs one photon, $P_\mathbf{exc}$ is the probability that the electron impact excites a triplet He$^*$ and $P_\mathbf{recomb}$ is the probability that the scattered electron recombines with the ion.
The simulated ICD rate plotted as blue dashed lines in Fig.~4 in the main text is thus obtained by normalizing the signal $S_\mathrm{ICD(1h\nu)}$ in Eq. (5) to $P_\mathrm{1} \times P_\mathrm{photo}$ where $P_\mathrm{photo}$ is the probability that the electron leaves the droplet without undergoing inelastic scattering,
\begin{equation}
N_\mathrm{ICD(1h\nu)} \approx \frac{P_\mathrm{exc}\times P_\mathrm{recomb}}{P_\mathrm{photo}}.
\end{equation}

The simulated number of ICD electrons plotted in black in Fig.~5 in the main text is obtained by normalizing the signal $S_\mathrm{ICD(1h\nu)}$ in Eq. (5) to $P_\mathrm{1}$,
\begin{equation}
N_\mathrm{ICD(1h\nu)} \approx \frac{P_\mathrm{1} \times P_\mathrm{exc} \times P_\mathrm{recomb}}{P_\mathrm{1}}  = P_\mathrm{exc} \times P_\mathrm{recomb}.
\end{equation}

 In the simulation of the ICD rate described in Eq. (6) we considered that all He$^*$ pairs produced by electron impact excitation and recombination undergo ICD. Thus, the ICD probability is assumed to be equal to 1 which probably is unrealistic. He$^*$ atoms emerging to the droplet surface can separately detach from the droplets; possibly, ICD happens after the droplet has moved out of the detection region of the spectrometer ($\gtrsim10\mu$s). This may explain the scaling factor 0.3 that is needed to scale down the simulated ICD rate plotted as blue dashed lines in Fig.~4. The simulation of the ICD channel where two He$^*$'s are created by two recombination events after one photoionization followed by one electron impact ionization (orange dashed lines in Fig.~4 in the main text) likely is quite inaccurate because the mutual Coulomb repulsion of both the two electrons and the two ions is neglected in the simulations; The Coulomb repulsion likely hinders efficient $e$-He$^+$ recombination and ICD. This may explain the need of scaling this contribution down by a factor 0.1. Note that the migration of He$^*$'s through the droplet and their decay by ICD is not explicitly modeled in the simulation. Neither is the formation of void bubbles around the electrons taken into account, which likely results in underestimation of the $e$-He$^+$ recombination rate in the simulation.
\\\\
ICD in He droplets containing two or more He$^*$ excitations has previously been studied using intense XUV pulses generated by an FEL~\cite{Ovcharenko:2020,LaForge:2021}. There, the absorption of two or more photons by one droplet was the dominant excitation scheme. In the present synchrotron experiments two-photon absorption is still present although the XUV peak intensity is lower by 12 orders of magnitude and the number of photons per pulse is lower by 8 orders of magnitude. Owing to the large absorption cross section of large He droplets $R\gtrsim20$~nm ($\langle N\rangle\gtrsim10^6$ He atoms), two or more photons can be absorbed on the strong 1s2p\,$^1P$ droplet resonance, for which we estimated an absorption cross section per He atom of 25~Mb~\cite{BuchtaJCP:2013,Ovcharenko:2020}. However, at higher photon energies $h\nu\gtrsim45$~eV the absorption cross section is $\lesssim 2$~Mb and we argued that ICD is predominantly induced by one-photon ionization followed by inelastic scattering and $e$-He$^+$ recombination. Here we support this conjecture by simple estimates. The probability that a He nanodroplet of radius $R= 75$~nm ($\langle N\rangle\approx10^{8}$ He atoms) resonantly absorbs one photon ($P_\mathrm{1}$ ) or two photons ($P_\mathrm{2}$) can be estimated by~\cite{BuchtaJCP:2013}
\begin{equation}
P_\mathrm{1} = N_\mathrm{p} \Phi T_\mathrm{p}\langle N\rangle \sigma, \hspace{0.6 cm}  P_\mathrm{2}\approx P_\mathrm{1}^2.
\end{equation}
Here, $\Phi$ is the time-averaged photon flux at given photon energy $h\nu$, $T_p = 2~$ns is the repetition period of the light pulses, $N_p\approx10^3$ is the number of pulses each He droplet has the chance to interact with during its flight across the interaction region, and $\sigma$ is the absorption cross section of He nanodroplets. At $h\nu$= 21.6~eV, the photon flux is $\Phi\approx$ 3$\times$10$^{13}$~s$^{-1}$cm$^{-2}$ and $\sigma$= 25~Mb. This leads to an estimated $P_\mathrm{1}\approx0.15$ and $P_\mathrm{2}\approx0.02$. Thus, a significant fraction of the droplets is two-photon excited, in agreement with the observed ICD electron signals (cyan lines in Fig.~2 in the main text). At $h\nu= 46.4$~eV, however, $\Phi\approx7\times10^{13}$~s$^{-1}$cm$^{-2}$ and $\sigma=2.3$~Mb~\cite{Samson:2002}. This yields $P_\mathrm{1}\approx0.03$ and $P_\mathrm{2}\approx0.001$. As $e$-He$^+$ recombination is highly effective in such large droplets according to our trajectory simulations, one-photon absorption is the dominant process for electron-impact-induced ICD. Higher-order processes such as two-photon ionization followed by $e$-He impact excitation and $e$-He$^+$ recombination are neglected in the simulation.

The number of ICD events due to two-photon photoionization followed by electron-impact excitation can be determined from the number of such one-photon trajectories weighted by the probability of two-photon absorption. The ICD signal per droplet and following two-photon ionization can therefore be expressed by
\begin{equation}
S_\mathrm{ICD(2h\nu)} \approx (P_\mathrm{1} \times P_\mathrm{exc})^2.
\end{equation}

The simulated ICD rate due to two-photon photoionization and plotted by cyan dashed lines in Fig.~4 in the main text is thus obtained by normalizing the signal $S_\mathrm{ICD(2h\nu)}$ in Eq. (9) to $P_\mathrm{1} \times P_\mathrm{photo}$,
\begin{equation}
N_\mathrm{ICD(2h\nu)} \approx \frac{P_\mathrm{exc}^2}{P_\mathrm{photo}} \times P_\mathrm{1}.
\end{equation}

The simulated number of ICD electrons due to two-photon photoionization and plotted by cyan solid line in Fig.~5 in the main text is obtained by normalizing the signal $S_\mathrm{ICD(2h\nu)}$ in Eq. (9) to $P_\mathrm{1}$,
\begin{equation}
N_\mathrm{ICD(2h\nu)} \approx \frac{(P_\mathrm{1} \times P_\mathrm{exc})^2}{P_\mathrm{1}} = P_\mathrm{exc}^2 \times P_\mathrm{1}.
\end{equation}

\section{Estimates of the kinematics and kinetics of electron-helium elastic scattering}
At first thought it may appear surprising that electrons can loose a substantial amount of their kinetic energy by purely elastic scattering at He atoms, given the large mismatch in mass. Here we present simple estimates of the kinematics of $e$-He collisions. The maximum energy transfer in an elastic collision of an electron with a He atoms occurs in a direct head-on collision. Then, the energy loss per collision is~\cite{Henne:1998}, 
\begin{equation}
\Delta E=E-E'=\frac{m}{2}v^2 - \frac{m}{2}v'^2 = \frac{m}{2}v^2\left[\left(\frac{M-m}{M+m}\right)^2-1\right]\approx 4 E\frac{m}{M}.
\end{equation}
 Here, $v$ and $v'$ is the electron velocity before and after the collision, respectively, and $m=m_e=1$, $M=m_\mathrm{He}\approx 8000$. Thus, for $E=10$~eV, this yields an energy loss per collision of up to $\Delta E=5$~meV.
	
The corresponding factor by which the electron energy is altered in a collision is
\begin{equation}
x=\frac{E_e'}{E_e}=\frac{v'^2}{v^2}=\left(\frac{M-m}{M+m}\right)^2\approx 1-4\frac{m}{M}=99.95\,\%,
\end{equation}
see \footnote{\url{https://en.wikipedia.org/wiki/Elastic_collision}}. 

Thus, it takes $k=1000$ collisions to reduce the electron energy by a factor
\begin{equation}
x^k \approx \left(1-4\frac{m}{M}\right)^k\approx 1-4k\frac{m}{M}\approx\frac{1}{2}.
\end{equation}
This number of collisions commonly occurs in large He droplets $R\gtrsim20$~nm containing $\langle N\rangle\gtrsim10^6$ He atoms.

Assuming an elastic scattering cross section for $e$-He atom collisions of $\sigma_\mathrm{elast}=5$~\AA{}$^2$~\cite{golden1965absolute}, the mean free path is
\begin{equation}
\ell_\mathrm{elast}=\frac{1}{\sigma_\mathrm{elast}  n_\mathrm{He}}\approx 0.9~\mathrm{nm},
\end{equation}
where $n_\mathrm{He}=22$\,nm$^{-3}$ is the density of He atoms in the bulk of He droplets. Thus, for the electron to scatter $k=1000$ times (reduction of its energy by 1/2), the He droplet needs to have the size
\begin{equation}
R\sim \sqrt{k\ell_\mathrm{elast}} \approx 30~\mathrm{nm},
\end{equation}
when assuming a random-walk-like motion of the electron. This result roughly agrees with the size of the droplets where we see that recombination is effective, and also with the average $e$-He$^+$ distance previously reported for bulk liquid He~\cite{McKinsey:2003}. Furthermore, the electron energy loss is enhanced by a rising total elastic scattering cross section and by the increasingly backward-peaked differential scattering cross section at low collision energy~\cite{Adibzadeh:2005}, which is not taken into account in this simple estimate.

The total inelastic scattering cross section around $h\nu=46$~eV is $\sigma_\mathrm{inelast}\approx 0.1$~\AA{}$^2$~\cite{Ralchenko:2008}. So the mean free path is 
\begin{equation}
\ell_\mathrm{inelast}=\frac{1}{\sigma_\mathrm{inelast} n_\mathrm{He}}\approx 46~\mathrm{nm},
\end{equation}
which yields a mean collision time for electrons at $E_e=21.4$~eV of
\begin{equation}
t_\mathrm{inelast}=\frac{\ell_\mathrm{inelast}}{\sqrt{2E_e/m_e}}\approx 15\,\mathrm{fs}, 
\end{equation}
in good agreement with our numerical simulation, \textit{cf.} Fig.~3 c) of the main text. 
	
The transfer of nearly 10~eV of electron energy to the He droplet by multiple elastic scattering naturally leads to massive evaporation of He atoms. However, given the large size of the droplets needed for stopping the electron ($N_\mathrm{He}\gtrsim 10^7$ He atoms), this is still a small fraction of the total vaporization energy of the droplet $N_\mathrm{He}\times0.62 $~meV~$\approx6$~keV.
  
%